\documentclass[twocolumn,prb,showpacs]{revtex4}

\usepackage{graphicx}
\usepackage{rotating}
\usepackage{amsmath}
\usepackage{amsfonts}
\usepackage{amssymb}
\usepackage{enumerate}
\usepackage{longtable}
\usepackage{dcolumn}

\usepackage{bm}

\begin{document}

\newcommand{\be}{\begin{equation}}
\newcommand{\ee}{\end{equation}}
\newcommand{\bn}{\begin{eqnarray}}
\newcommand{\en}{\end{eqnarray}}

\title{Quantum Critical Phase and Lifshitz Transition in an Extended
Periodic Anderson Model  }

\author{M. S. Laad$^{1}$, S. Koley$^{2}$ and A Taraphder$^{2}$}

\affiliation{$^{1}$Institut f\"ur Theo. Physik, RWTH Aachen University,
Aachen 52056, Germany.\\$^{2}$ Department of Physics and Centre for
Theoretical Studies, Indian Institute of Technology, Kharagpur 721302,
India.}

\date{\today}

\begin{abstract}
We study the quantum phase transition in $f$-electron systems as a quantum
Lifshitz transition driven by selective Mott localization in a realistic
extended Anderson lattice model.  Using DMFT, we
find that a quantum critical {\it phase} with anomalous $\omega/T$ scaling
separates a heavy Landau-Fermi liquid from ordered phase(s).  Fermi surface
reconstruction occurs via the interplay between, and penetration of the Green
function zeros to the poles, leading to violation of Luttinger's
theorem in the selective-Mott phase .  We show how
this naturally leads to scale-invariant responses in transport.  Our
work is represents a specific (DMFT)
realization of the hidden-FL and FL$^{*}$ theories, and holds promise
for study of ``strange'' metal phases in
quantum matter.

\end{abstract}


\maketitle

\section{INTRODUCTION}
 Magnetic quantum phase transitions (MQPT) from antiferromagnetically ordered
metals to heavy landau-fermi liquids (HLFL) are believed to occur in a
large class of $f$-electron compounds as an appropriate external
parameter is tuned~\cite{gil,steglich}.  Hertz-Moriya-Millis
(HMM)~\cite{Hertz} theory ``breaks down'' for a
 sub-class of these systems: the whole Fermi surface (FS), or a subset
of FS sheets in multi-band cases,
is destabilized (becomes ``hot''), and an abrupt Fermi surface
reconstruction (FSR), together with anomalous scale-invariant
power-law responses,
accompanies the QPT.  These features are also found in near-optimally
doped cuprates and favor a quasi-local picture
based on an selective
Mott transition (SMT) in terms of either a Kondo-RKKY lattice
model~\cite{si}, Hubbard~\cite{haule} or periodic
Anderson lattice model~\cite{senthil,pepin}.  On a fundamental side,
these findings have spawned intense theoretical
interest in view of the
fact that the specific scale-invariant responses they exhibit are
fundamentally at odds with the Hertz-Moriya-Millis
(HMM) approaches to quantum criticality: the scale invariant
power-laws are only observed as a function of energy,
and the responses have very weak momentum (${\bf k}$) dependence.

 These findings have spurred intense theoretical activity on various
fronts.  Slave-boson-Hartree-Fock-plus gauge
field fluctuation approaches have extended earlier studies for the
$t-J$ model~\cite{lee} to the
Kondo-RKKY~\cite{senthil} as well as extended periodic
Anderson~\cite{pepin} models.  An attractive proposal in
this context is the idea of the FL$^{*}$ metal, which is argued to
result from a decoupling of local moments from
conduction electrons at the QPT.  This FL$^{*}$ state is exotic, with
fractionalised excitations.
 Within extended-DMFT~\cite{si}, such physics (loss of Fermi Liquid
coherence) occurs right at the QCP associated
with onset of magnetic order.  On the other hand, the FL$^{*}$
proposal {\it et al.}~\cite{senthil} is not tied
down to this feature; it only {\it requires} short-range fluctuating
magnetism, but no AF order.  A recent proposal
of Anderson~\cite{pwa-KLM} argues that the Kondo lattice must have a
``massively non-Fermi liquid'' phase, akin to
the ``hidden-FL''~\cite{pwa}, where strong correlations would
completely deplete the Landau quasiparticle pole in
the fermionic Green's functions in favor of a branch-cut structure
(i.e, an incoherent continuum) in the infra-red.
Extant dynamical mean-field theoretic (DMFT) works~\cite{kotliar} have
studied the issue of the FSR in detail within
cellular-DMFT.  However, the full set of issues raised by the FL$^{*}$
and the hidden-FL ideas have not been
considered (within a fermionic modelling), as far as we are aware.

 In the fermionic context, it is also important to inquire whether the
features discussed above, and in particular,
the $\omega/T$-scaling and FSR, survive additional
{\it realistic} features: real rare-earth systems of interest have two
features neglected in the minimal
Anderson lattice model: finite $f$-bandwidth {\it and} ${\bf
k}$-dependent hybridization~\cite{pepin}: both these
features are clearly visible in first-principles LDA one-electron band
structure calculations~\cite{gertrud}.
It is not a priori clear, especially within DMFT approaches, how these
additional delocalising features might affect
the local critical behavior hitherto considered.  For example, it is
well-known that one-electron hybridisation is
very often a relevant perturbation, and, in DMFT, can cut off
infra-red local singular behavior at low energies
under certain conditions, for e.g, in the spinless Falicov-Kimball
model, a finite interband hybridisation reverts the metallic
(non-symmetry broken) state to a correlated LFL.  Thus, studying the
stability vis-a-vis fragility of the ``strange'' metal under more
realistic conditions is indispensable for
understanding the ``strange'' metal in real systems~\cite{gil,steglich,kotliar}.

\section{Hamiltonian}

Hence, we investigate a more ``realistic'' {\it extended}-PAM with finite
$t_{ff}$, a ${\bf k}$-dependent hybridization, $V({\bf k})$, and finite
$f-c$ coulomb interaction.  All these features are relevant to a more
``realistic'' model for the systems of interest:
 the $f-c$ coulomb interaction, for instance, has long been known as a
key factor even for the well-known $\alpha-\gamma$ transition in
elemental Cerium~\cite{allen}.  We thus expect that it should have
some relevance to real Ce- or
Yb-based compounds.

\noindent In view of the above, we start with the Hamiltonian
$H=H_{0}+H_{1}$ with

$$H_{0}=-t_{f}\sum_{<i,j>,\sigma}f_{i\sigma}^{\dag}f_{j\sigma} -
t_{c}\sum_{<i,j>,\sigma}c_{i\sigma}^{\dag}c_{j\sigma} $$
$$+ V_{fc}\sum_{<i,j>,\sigma}(f_{i\sigma}^{\dag}c_{j\sigma}+h.c)$$\\
and the local part is

$$H_{1}= U_{ff}\sum_{i}n_{if\uparrow}n_{if\downarrow} +
U_{fc}\sum_{i,\sigma,\sigma'}n_{if\sigma}n_{ic\sigma'}
+ \epsilon_{f}\sum_{i}n_{fi} $$
$$+ V\sum_{i}(f_{i\sigma}^{\dag}c_{i\sigma}+h.c)$$

 For $U_{fc}=0$, emergence of a heavy LFL within DMFT is well
known~\cite{jarrell}.
 Taking the $c$-band centered around $E=0$, we investigate the fate of
DMFT Kondo scale in presence of strong, quantum fluctuations
of the $f$ occupation, caused by the competition between
$V_{fc}$ and incoherence, driven by $U_{fc}$.

\begin{figure}
\centering

{\includegraphics[angle=270,width=\columnwidth]{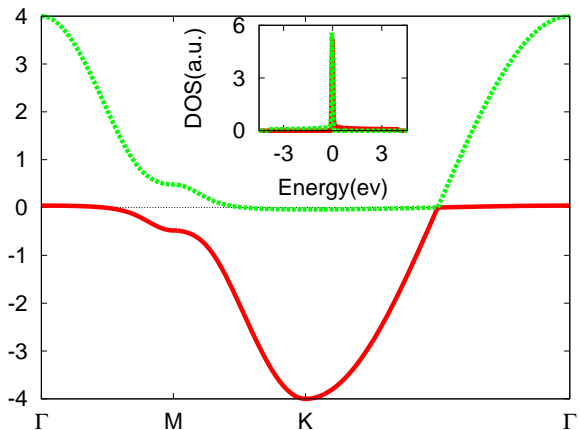}
}

\caption{(Color online) Unperturbed band structure including $d$-wave
hybridization
for the two-band model in the text.
}

\label{fig1}
\end{figure}

\section{DMFT Results and Selective-Mott Transition}

We solve $H$ using the multi-band iterated-perturbation theory (IPT)
at arbitrary $T$.  Though not numerically exact, it is a fast solver for
multi-band cases at arbitrary $T$ and band-filling, recovers known LFL metal for
$U_{fc}=0$~\cite{jarrell}, and gives good quantitative accord with one- and
two-particle spectral responses for a host of multi-orbital strongly correlated
systems~\cite{laad}.  We follow the general DMFT approach~\cite{gkkr}, and focus
only on symmetry-unbroken states hereafter.

 The case $\epsilon_{f}=0,V=0$ is analytically soluble in DMFT for a special
choice of $V_{fc}^{(c}=\sqrt{t_{f}t_{c}}$.
 At $V_{fc}^{(c)}$ this becomes
the $S=1/2$ Falicov-Kimball model (FKM) with $a,b$ fermionic combinations~
\cite{epl} and a rigorous
{\it local} U(1) symmetry associated with $[n_{i,a},H]=0$ at {\it
each} site.  Within DMFT, $G_{bb}(\omega)$
shows the upper- and lower-Hubbard bands {\it without} the
renormalised (LFL) lattice Kondo resonance.  Concomitantly,
$G_{aa}(i\omega_{n})\simeq i|\omega_{n}|^{-(1-\eta)}$sgn$\omega_{n}$,
and $\alpha=(\delta/\pi)^{2}$ with
$\delta=(2/\pi)$tan$^{-1}(\pi\rho_{a}(0)U_{fc}/2)$ is the Anderson
orthogonality-catastrophe
(OC) exponent, rigorously true for the FKM in DMFT~\cite{gkkr}.
Correspondingly, the local excitonic susceptibility,
$\chi_{ab}(\omega)=\langle
a_{i\sigma}^{\dag}b_{i\sigma};b_{i\sigma}^{\dag}a_{i\sigma}\rangle
\simeq -i|\omega_{n}|^{-(2\eta-\eta^{2})}$, is also infra-red
singular.
This is an {\it analytic} demonstration of the intimate link between
selective Mott physics and (conformally
invariant) local QCP in DMFT.  In terms of $f,c$ fermions, the local\
$G_{ff}$ and $\chi_{fc}(\omega)$ are infra-red singular for a {\it
finite}
$V_{fc}^{(1)}=\sqrt{t_{f}t_{c}}$.  A finite $\delta
V_{fc}=(V_{fc}-V_{fc}^{(c)})$ is known to lead to a finite angular
momentum ($l=2$) LFL for $U_{fc}=0$~\cite{sentil}.  Thus, $\delta
V_{fc}=0=\epsilon_{f}=0=V$ is a local QCP in the
PAM with a {\it finite} non-local hybridization, separating two
correlated LFL phases: this is a selective Mott
metal, and the LFL is destroyed by the OC~\cite{pwa}.
 Though known for the FKM~\cite{si}, such a local QCP has not been
found previously for the PAM with a finite $V_{fc}({\bf k})$ within
DMFT.
 Since $U_{fc}>0$ inhibits lattice Kondo (LFL)
coherence in the usual PAM, one expects the local QCP found above to extend into
a local non-LFL {\it phase} (see below) between critical values
$U_{fc}^{(1)}<U_{fc}<U_{fc}^{(2)}$, going over to a
Mott insulator for $U_{fc}>U_{fc}^{(2)}$ and a heavy LFL for
 $U_{fc}<U_{fc}^{(1)}$.  Below, we show that such a local QC phase is intimately
tied to an orbital-selective Mott localization of the $a$-fermions, and is a
manifestation of the {\it fractionalized} Fermi liquid~\cite{senthil,zaanen}
within DMFT.

 We start with two ($c,f$) bands in $D=2$,
modelled by a simplest nearest-neighbor tight-binding dispersion:
$\epsilon_{f,c}(k)=2t_{f,c}($cos$k_{x}+$cos$k_{y})$.  Motivated by
real systems~\cite{gil,steglich}, we choose
$V_{fc}(k)=V_{fc}($cos$k_{x}-$cos$k_{y})$ to have
a $d$-wave form factor, and also keep a
small $\epsilon_{f}$ (measured from $E_{F}(=0)$).  In Fig.~\ref{fig1},
we show the unperturbed band structure and local
density-of-states (DOS) for selected values of
parameters ($t_{f},t_{c},V_{fc},V$ = 1.0, 0.01, 0.24, 0): this is
 an input DOS in the DMFT.  For concreteness, we choose
$U_{ff}=U_{cc}=3.0$~eV (intra-orbital Hubbard
$U$), vary the inter-orbital Hubbard interaction $U_{fc}$, and focus
on correlation-induced spectral changes in the DMFT propagators,
$G_{a,b}(\omega)$ and self-energies, $\Sigma_{a,b}(\omega)$.

\begin{figure}
{(a)}
{\includegraphics[angle=270,width=\columnwidth]{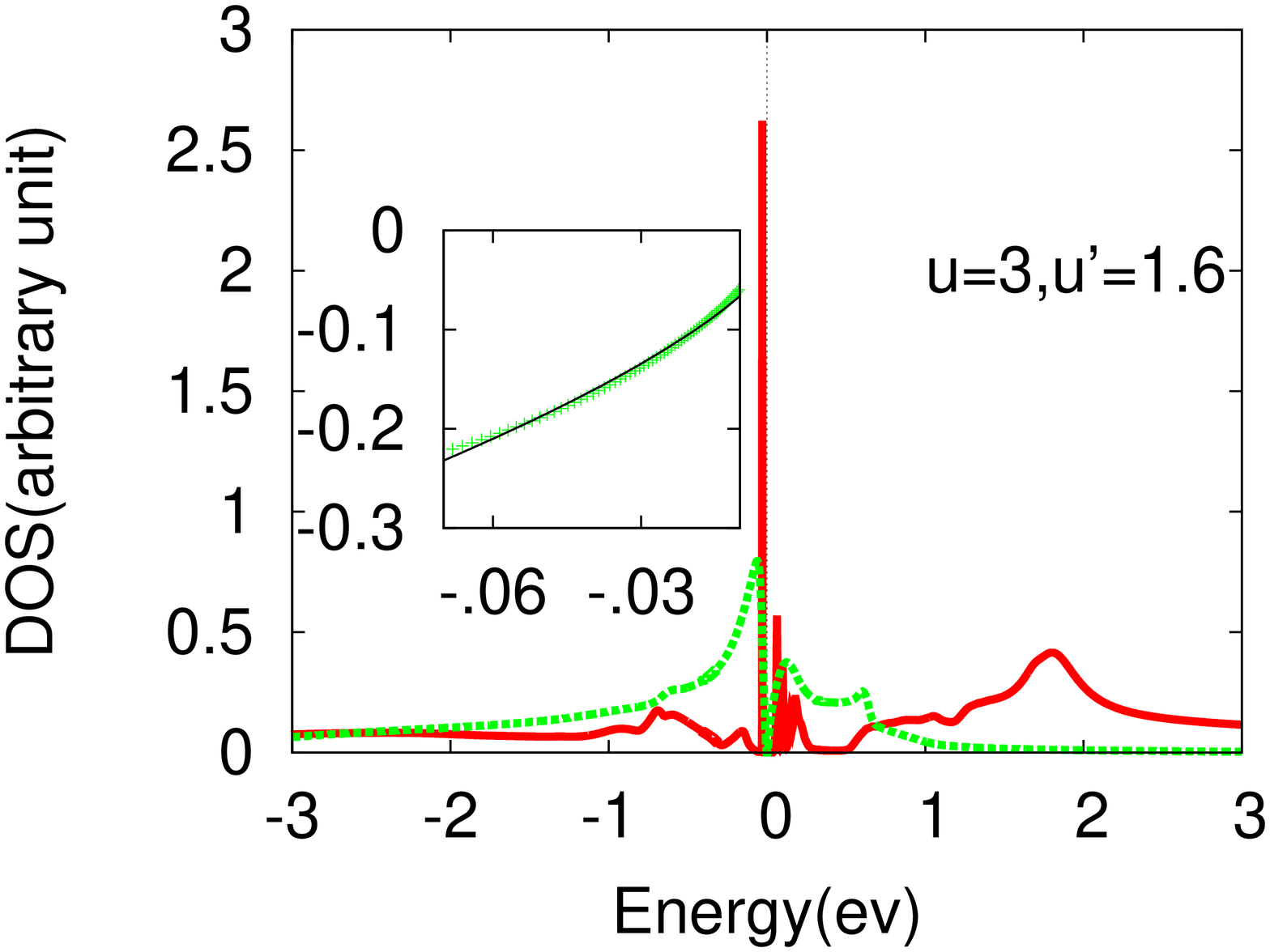}
}
{(b)}
{\includegraphics[angle=270,width=\columnwidth]{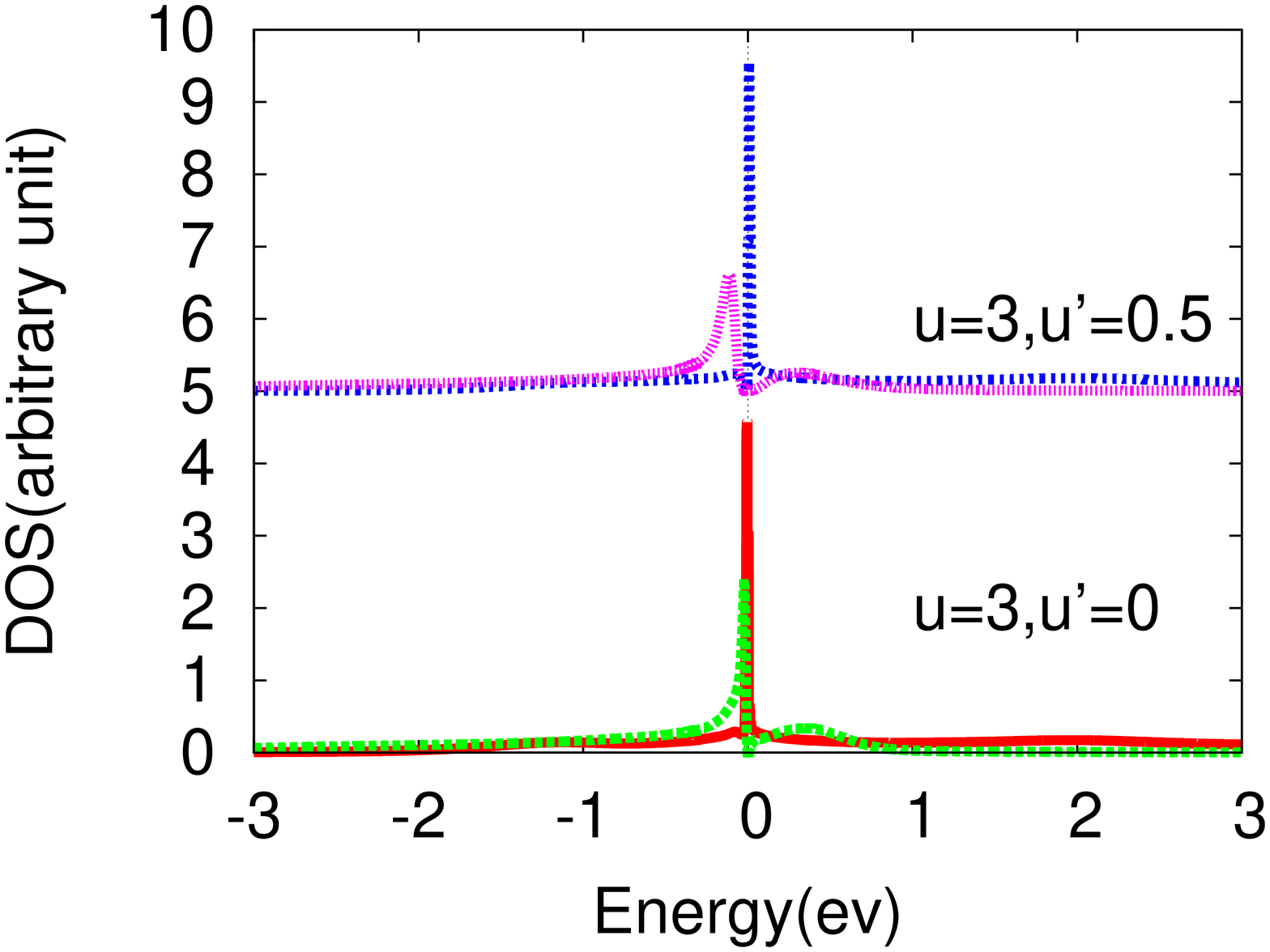}
}

\caption{DMFT DOS for the two band model in the QC metal phase
($U_{fc}>U_{fc}^{(1)}$,
 see text, green and red refer to `b', `a' bands) clearly showing
power-law fall-off (inset: the black line is the power law fit to
the green line, Im$\Sigma_b(\omega)$, in a small range below fermi level) 
with energy in the infra-red,
along with the OS Mott transition (upper panel).
Lower panel shows DMFT DOS for the heavy LFL state for $U_{fc}=u^{'}=0.0,0.5$~eV
 and $U_{ff}=U_{cc}=u=3.0$~eV
(red
and blue refer to `a' band; the green and purple refer to `b' band).}

\label{fig2}
\end{figure}

\begin{figure}
{(a)}
{\includegraphics[angle=270,width=\columnwidth]{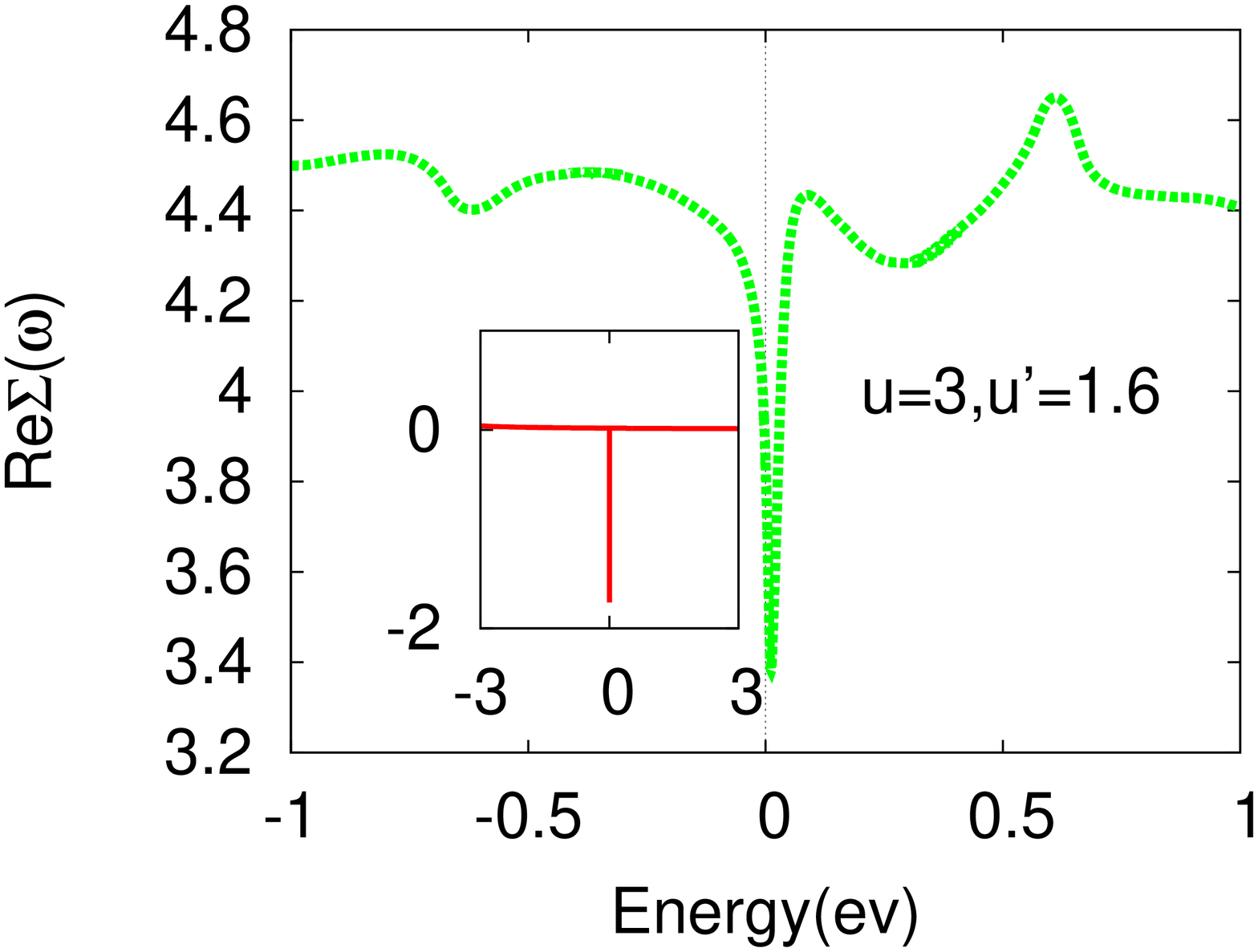}
}
{(b)}
{\includegraphics[angle=270,width=\columnwidth]{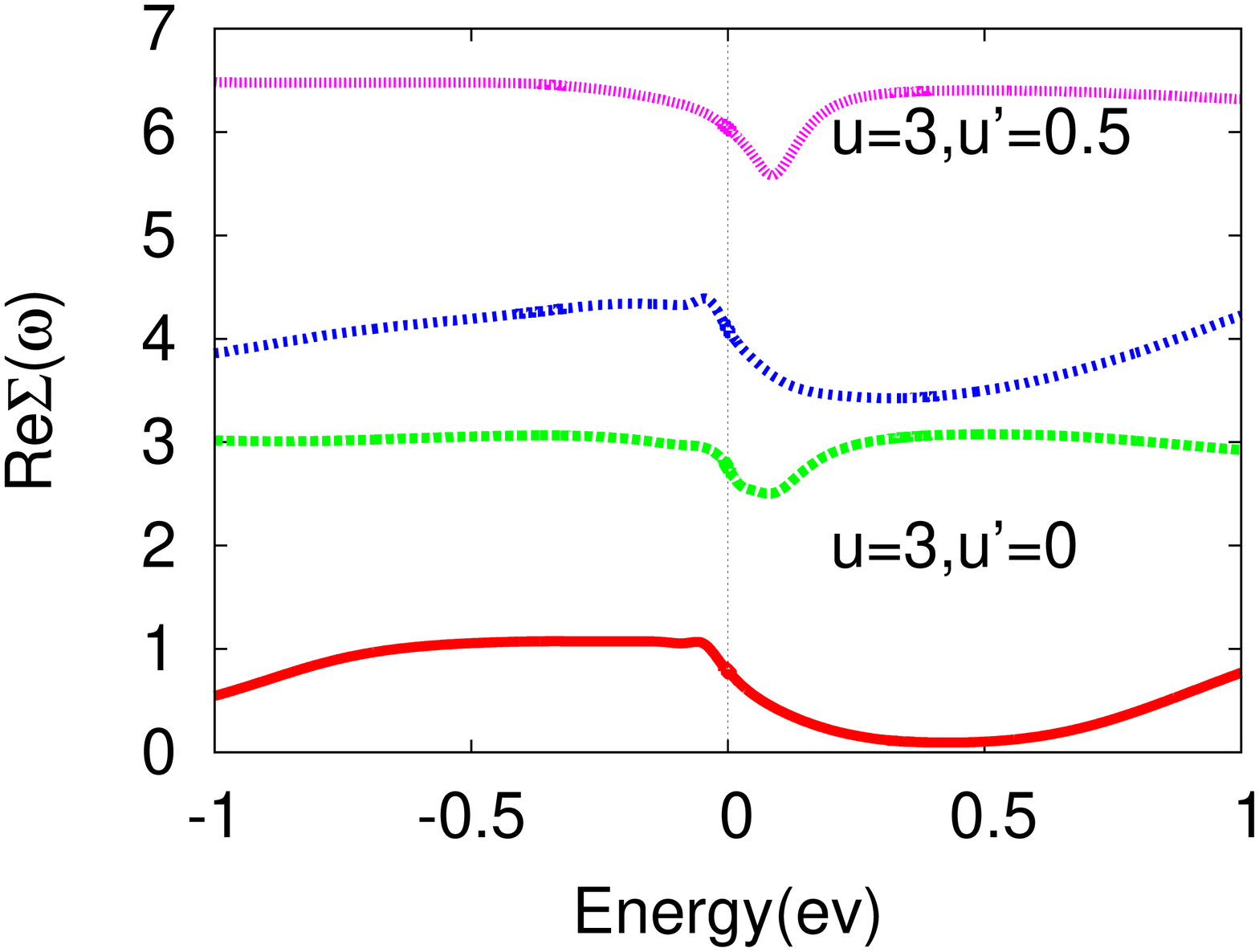}
}
\caption{(Color online) Real part of self-energies for the critical
(non-LFL) metal (upper panel) and for the heavy LFL (lower panel)
within DMFT.  Clear
mass divergence and insulating behavior (inset of upper panel, x-axis and
y-axis to be multiplied by 0.001 and 1000, respectively) and correlated
LFL behavior (lower panel) are manifest. Red (in inset, over a narrow range)
and green in upper panel represent
the two bands. In the lower panel red and blue stand for band `a' while
green and purple stand for band `b' at two different $u, \, u^{'}$ values.
}

\label{fig3}
\end{figure}

Fig.~\ref{fig2} shows the DMFT spectral functions for the
$a,b$~\cite{epl} bands.
Several remarkable features stand out: (i) the $a$-band is Mott-split,
(ii) the $b$-band DOS clearly shows
infra-red singular behavior, with a power-law fall-off in energy upto
rather high energy, $G_{bb}(\omega)\simeq
c_{1}\theta(\omega)|\omega|^{-(1-\eta)}+c_{2}\theta(-\omega)|\omega|^{-(1-\eta)}$
with $\eta\simeq 2/3$ (inset, upper panel, Fig.2), and a pronounced
low-energy asymmetry.  Correspondingly, at long (imaginary) times,
$G_{bb}(\tau)\simeq \tau^{-(2-\eta)}$ instead of
$G_{bb}(\tau)\simeq 1/\tau^{2}$ for any LFL.  Now, extending this to
finite $T$, Fourier transformating to
Matsubara frequencies, followed by an analytic continuation of
imaginary frequencies onto the real energy axis,
 following similar analysis for Luttinger liquids~\cite{sensa} gives

\be
G_{bb}(\omega,T)=c_{1}T^{-(1-\eta)}F_{0}(\omega/T)
\ee
where $F_{0}(\omega/T)$ is a universal scaling function.  The
one-fermion spectral function at finite $T$ thus
exhibits $\omega/T$ scaling with an anomalous exponent in the
infra-red as a consequence of the Anderson OC, as
in the ``simpler'' toy model above.
 (iii) concomitantly, Im$\Sigma_{b}(\omega) \simeq |\omega|^{1-\eta}$,
while Im$\Sigma_{a}(\omega)$ clearly
shows a pole at $\omega=0$ characteristic of a Mott insulator, shown
in Fig.~\ref{fig4} and
(iv) Re$\Sigma_{b}(\omega)$, shown in upper panel of Fig.~\ref{fig3}
clearly reveals
vanishing LFL quasiparticle residue,
$Z=[1-(\partial\Sigma_{b}/\partial\omega)_{\omega=0}]^{-1}=0$, on the
FS,
implying a divergent effective mass, $m_{b}^{*}/m_{b}=1/Z$.
This singular branch-cut form of $G_{bb}$ is a fundamentally non-perturbative
manifestation of selective (Mott)
localization, and {\it cannot} be obtained in weak-coupling perturbation theory.
This OC-induced local critical phase also implies singular hybridization
fluctuations: the ``excitonic'' susceptibility also shows an infrared
singularity,
$\chi_{ab}(\omega)\simeq \theta(\omega)|\omega|^{-(2\eta-\eta^{2})}$, leading to
soft local {\it valence} fluctuations~\cite{miyake}.  This particular
aspect might
be of interest in connection with the non-LFL behavior of the recently
found $\beta-YbAl_{4}$~\cite{gill}

\section{Connection to Hidden-FL and FL$^{*}$ Metals}

 Remarkably, this selective metal is a singular Fermi fluid showing
most of the central features of the hidden
FL~\cite{pwa} and fractionalized FL (FL$^{*}$)~\cite{senthil}: (i) implies that
$a$-fermion states do not contribute to the
reconstructed FS, and Luttinger's theorem does not hold.  This is
because the pole in
 Im$\Sigma_{a}(\omega=0)$ gives a {\it finite}
$\int_{-\infty}^{+\infty}(d\omega/2\pi)\Sigma_{a}(\omega)(\partial
G_{a}(\omega)/\partial\omega)=-$sign$(\epsilon_{f})/2$, using the effective
atomic limit for $G_{a},\Sigma_{a}$~\cite{rosch}, violating
Luttinger's theorem.  Though Rosch
analysed this case for the Mott insulator within the Hubbard-I
approximation, the Mott-localisation of the $a$-band
states within DMFT allows us to use the same argument without
modification for our metallic case.
 In reality, the unquenched spin degrees of freedom from the now
localised $a$-fermions are now described by an
effective spin-1/2 Heisenberg model at low energy.  The superexchange
interaction between these spins will now be
mediated by the metallic $b$-band carriers, as in an RKKY scenario.
An important difference with the usual RKKY case
is that the ``conduction fermion'' sea now corresponds to carriers
which are {\it not} Landau quasiparticles, thanks
to the branch-cut structure of $G_{bb}(\omega)$ found above.  This has
remarkable consequences when we consider the
dynamical spin susceptibility later (see below).

 As long as these do not give AF long-range order, our results imply
that the spin
excitations arising from such a Heisenberg model in the $a$-fermion sector are
distinct and asymptotically decoupled from the $b$-fermion states making up
the new FS.  The $b$-spectral function
shows power-law singular behavior of the $(0+1)$-D CFT,
Im$G_{b}(\omega)=\theta(\omega)|\omega|^{-(1-\eta)}$,
upto high energy and a pronounced particle-hole (p-h) asymmetry.  In
hidden-FL theory, LFL
behavior is destroyed by the OC arising from blocking of recoil
processes~\cite{pwa}, while an
OSMT is a key requirement for the FL$^{*}$ state.  At least at DMFT
level, we see that an OSMT
and the OC are inseparable phenomena.  Generation of Mott localised
$a$-fermion states in the OSMT directly implies
strong scattering between localised ($a$) and ``itinerant'' ($b$)
fermions: in DMFT, the corresponding impurity
model is precisely the X-ray edge problem.  The infra-red singular
branch-cut form of the DMFT spectra is then a
consequence of the lattice ``orthogonality catastrophe'', very similar
to the hidden-FL theory.
 It also follows that the local QCP is intimately tied
to destruction (via $U_{fc}$) of the lattice Kondo scale: once the
$a$-states are Mott-localized,
 the $a-b$ hybridization is {\it irrelevant} and Kondo screening cannot occur.
 From (iii), we do find a FS, since Im$\Sigma_{b}(\omega=0)=0$, but
{\it without} electron-like quasiparticles
($Z=Z_{b}=0$).  In contrast to the analytic $S=1/2$ FKM argument
above, this local critical metal survives a finite range of (see
below) $U_{fc}^{(1)}=0.6$~eV$<U_{fc}<U_{fc}^{(2)}$ (the ``strange''
metal goes over to a Mott insulator
for $U_{fc}>U_{fc}^{(2)}$, not shown) in the DMFT.  Thus, we find a
quantum critical
 phase, again in accord with the hidden FL and FL$^{*}$ theories.  This is a
new and novel aspect of our work: earlier DMFT works have focussed on
quantum critical {\it point(s)} at $T=0$, while we find a QC {\it
phase}.

\begin{figure}
{(a)}
{\includegraphics[angle=270,width=\columnwidth]{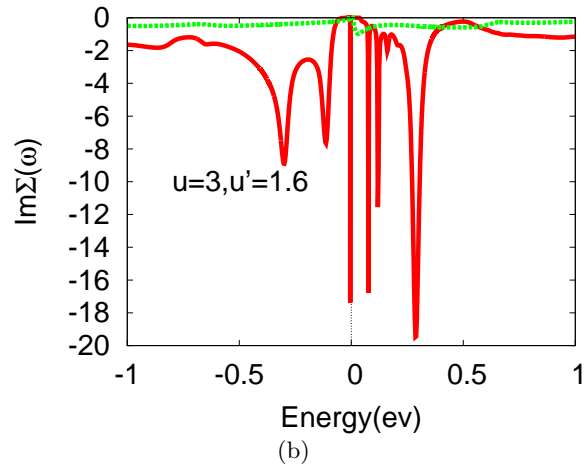}
}
{(b)}
{\includegraphics[angle=270,width=\columnwidth]{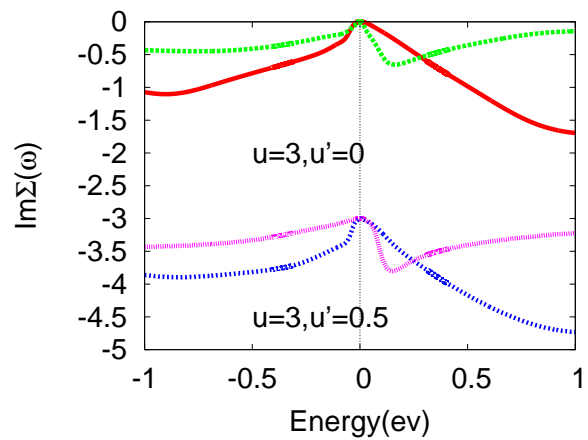}
}

\caption{(Color online) Imaginary parts of the self-energies for the critical
metal (upper panel) and the heavy LFL (lower panel) within DMFT.
Clear non-LFL behavior (green) and selective Mott (red) behavior is
seen (upper panel)
and low-energy heavy LFL behavior (lower panel) is visible ($u, \,
u^{'}$ as shown in figure, rest of the parameters same as in fig.1).
}
\label{fig4}
\end{figure}

 A heavy LFL metal, with correlation-induced Hubbard bands and a
low-energy pole in $G_{bb}(\omega)$, with a small
but finite $Z$ is already known for $U_{fc}=0$~\cite{jarrell}.
A small enough $U_{fc}$ cannot destabilize the lattice (Kondo) LFL
coherence scale, so this feature should survive
up to a critical $U_{fc}=U_{fc}^{(1)}$.  The lower panels in
Fig.~\ref{fig2}-Fig.~\ref{fig4} bear
this out, and show low-energy LFL features up to $U_{fc}=0.6$~eV
instead of OS-Mott-induced non-LFL
behavior.  Thus, heavy LFL metallic behavior obtains for
$0<U_{fc}<0.6$~eV, while the local ``strange'' metal
is found for $U_{fc}>0.6$~eV, as promised before.
LFL behavior is intimately tied to relevance (in the RG sense) of the
hybridization in the PAM, and, when this occurs,
 (coherent) $a$-spectral weight is pulled down to $E_{F}$, whence no
OS-Mott physics can exist in the LFL state.
Obviously, the Luttinger volume now encloses {\it both} $a$ and $b$-fermions.
Since an OS-Mott transition is generically first-order~\cite{kotliar},
a discontinuous jump in
the FS volume from $V_{FL}=[(2\pi)^{2}/2v_{0}][(n_{a}+n_{b})($mod$2)]$ to
$V_{FL^{*}}=[(2\pi)^{2}/2v_{0}][n_{b}($mod$2)]$ rigorously follows across
$U_{fc}^{(1)}$ (with $v_{0}$ being unit-cell volume).
In the QC phase, the {\it zeros} of $G_{a}(\omega)$ cross $\omega=0$ (pole in
Im$\Sigma_{a}(\omega)$ at $\omega=0$) for $U_{fc}\geq U_{fc}^{(1)}$, and thus,
remarkably, the OSMT with FSR is a correlation-dominated quantum
(topolgical) Lifshitz transition (QLT)~\cite{imada}.  In DMFT, the
whole $a$-FS sheet disappears at this QLT.  On the
other hand, in cellular-DMFT, parts of the FS sheet(s) in momentum
space would disappear at a {\it momentum}-selective
 Mott transition~\cite{civelli}.
The quantum Lifshitz character of the selective-Mott transition is
further vindicated by the divergence of the
effective mass ($m_{b}^{*}=m_{b}/Z_{b}=\infty$) in the
QC phase, which implies vanishing quasiparticle {\it stiffness}, as
expected at a Mottness-induced QLT.
We call this a quantum Lifshitz transition, as opposed to
a classical one, which occurs, for e.g, in classical frustrated spin
systems~\cite{lubensky}.  Further, it is also
very different from a band-Lifshitz transition in a LFL metal: the
$a$-band FS sheet vanishes across this
quantum transition due to
selective Mottness arising from strong electronic correlations.  In a
usual one-electron picture, this is a simple
disappearance of a subset of FS sheet(s) due to band-folding effects,
and involves a completely filled one-electron
band, rather than any Mott physics.
An important consequence of selective-Mottness, out of scope of the
``traditional'' Lifshitz transition, is
that the Luttinger sum rule acquires a new meaning.  Recall that, in
DMFT, both, $G_{aa}(k,\omega), G_{bb}(k,\omega)$
have the form

\be
G_{\gamma\gamma}(k,\omega)=\frac{1}{\omega-\epsilon_{\gamma}(k)-\Sigma_{\gamma}(\omega)-\frac{V_{\gamma\gamma'}^{2}(k)}{\omega-\epsilon_{\gamma'}(k)-\Sigma_{\gamma'}(\omega)}}
\ee
with $\gamma,\gamma'=a,b$.  In the LFL phase, i.e, as long as
$\Sigma_{a,b}(\omega)$ show usual non-singular behavior,
$G_{aa,bb}(k,0)$ can only change sign from positive to negative
through an infinity in the G's at a closed FS comprising
{\it both} $a$ and $b$ fermions.  In the
selective metal, however, the qualitatively new, singular nature of
the DMFT self-energies introduces qualitatively
new aspects.
First, since Im$\Sigma_{bb}(\omega=0)=0$, the $b$-FS sheet is still
well-defined, even though LFL quasiparticles
have become extinct due to the OC.  However, in the selective-Mott
phase, the sign change can also occur via a
surface of zeros of $G_{aa}(k,\omega)$.  This surface of zeros, which
is where the zeros of
$G_{aa}(k,\omega)$ penetrate what would have been poles in the LFL
phase, is entirely distinct from the $b$-FS sheet,
 and, following Yang {\it et al.}~\cite{rice}, we dub it a Luttinger
surface of zeros.   The FS in the selective
metal encloses the $b$-fermions, while the more generalised Luttinger
surface defined above encloses an area given by\ the {\it total}
fermion number.  Whether this Luttinger surface of zeros can be
related to the ``ghost Fermi
surface'' proposed recently~\cite{friedel} is an interesting question:
in our DMFT study, the short-wavelength spin
singlet fluctuations (arising from  Mott-localised $a$-fermions as an
effective intersite ``RKKY'' superexchange,
see below) will, however, couple to a {\it critical} $b$-fermion
charge fluctuation spectrum.  In
Mross {\it et al.}'s study, both, the spin-singlet $2k_{F}$ response
and the electrical charge density arise from the
 {\it same} single band.  Nonetheless, the formal analogy pointed out
here is suggestive, and might facilitate
further study of the Luttinger surface in ``strange'' metals.

 Finally, given that the usual Anderson lattice model (ALM) within
two-site cluster-DMFT~\cite{liebsch} can be
re-expressed as our EPAM in terms of the bonding and anti-bonding
cluster-centered fermionic combinations, similar
results, with proper re-interpretation, will follow for the ALM within
cluster-DMFT~\cite{kotliar}.  In this latter
case, the self-energy will, of course, explicitly depend on cluster
momenta, and, for a two-site cluster, the
cluster-DMFT band structure will be modified in a momentum-dependent
way at the momentum-selective Mott transition.
In this case, using the mapping from cellular-DMFT to an effective
cluster-centered multi-orbital DMFT, the analogous
possibility that  momentum-selectivity~\cite{kotliar} within
cellular-DMFT induces a
similar phenomenon of penetration of zeros to poles along certain {\it
directions} in ${\bf k}$-space reveals itself.
In underdoped HTSC cuprates, this  momentum-selectivity reveals itself
as the nodal-antinodal dichotomy and FS pockets
in dHvA studies~\cite{rice}.
Thus, whether a DMFT or cluster-DMFT approach should be used must be
decided by appeal to dHvA experiments.  If
the renormalised band dispersions and FS agree well with LDA, at least
in the heavy LFL phase, one can surmise that
this k-dependence is then sufficiently weak that DMFT is suffices.
If, on the other hand, sizable discrepancy
between dHvA results and LDA shows up, one must look at the associated
selective-Mott and critical phenomena within
cluster-DMFT.

 Analytic insight into the non-LFL to LFL ``transition'' at $U_{fc}^{(1)}$ is
obtained by bosonizing
 the impurity model of DMFT.  In the QC metal, the impurity model is that of
$b$-fermions scattered by a localized $a$-fermion via $U_{fc}$.  This
maps exactly onto the X-ray-edge problem,
which is bosonizable into a collection of radial (free) bosonic models
on a half-line~\cite{schotte},
$0\leq r\leq\infty$, i.e,
$S=\int_{0}^{\infty}dr[\partial_{\tau}\phi_{\alpha}(r))^{2}+(\partial_{r}\phi_{\alpha}(r))^{2}]$,
whence the singularities above can be anaytically derived.  Relevance
of the hybridization (in the DMFT, this is the LFL
phase, where $V_{fc}(k)$ is relevant) necessitates proper treatment
of non-adiabatic effects (associated with finite recoil of the
$a$-fermion) in the bosonization approach via the
lowest-order cluster
expansion~\cite{mh}.  Carrying out this procedure changes the singular
branch cut structure of $G_{bb}(t)$ above to
$G_{bb}(t)=-i\theta(t)exp[-i(-\epsilon_{f}+U_{fc}n_{a})t-F_{b}(t)]$ with

\be
F_{b}(t)=(U_{fc}/N)^{2}\sum_{k,k'}n_{ka}(1-n_{k'a})\frac{1-e^{i\Delta_{a,kk'}t}+i\Delta_{a,kk'}t}{\Delta_{a,kk'}^{2}}
\ee
 and $\Delta_{a,kk'}==\epsilon_{f}+t_{a,k-k'}-t_{a,k}+t_{a,k'}$.
The $a$-spectral function is now $A_{b}(\omega)=2\pi
Z_{a}\delta(\omega-\epsilon_{a})+A_{inc}(\omega)$, with a {\it finite}
LFL QP residue,
$Z_{b}=exp[-(U_{fc}/N)^{2}\sum_{k,k'}\frac{n_{ka}(1-n_{k'a})}{\Delta_{a,kk'}^{2}}]<<1$.
The $a$-spectral function also develops a similar (exponentially small) $Z_{a}$
for $U_{fc}<U_{fc}^{(1)}$.  For energies $\omega,k_{B}T
>$min$(Z_{a},Z_{b})E_{F}$,
 this heavy LFL {\it smoothly} crosses over to the QC phase.  Interestingly, all
these DMFT spectral features, including the exponentially small
$Z$ increasing {\it smoothly} across $U_{fc}^{(1)}$, closely resemble
those found in very recent AdS-CFT work~\cite{sachdev,zaanen}.

\section{Spin and Charge Fluctuations in the ``Strange'' Metal}

In this section, we start by observing that anomalous spin and charge
correlations have turned out to be defining
characteristics of the ``strange'' metal in cuprates and
$f$-electron-based systems.  As mentioned
in the introduction, several careful studies now show that these have
a unique form, namely very anomalous energy dependence, but
essentially no ${\bf k}$-dependence, that does not fit into the
Hertz-Moriya-Millis approach to quantum criticality.  In particular,

(i) the dynamical spin susceptibility, $\chi_{\sigma}({\bf
q},\omega,T)$ is essentially independent of ${\bf q}$,
but shows anomalous power-law fall-off in the infrared as a function
of $\omega$: the interesting finding is
that it scales with $\omega/T$, implying no intrinsic energy scale
(apart from the temperature itself) in the
``strange'' metal.  This is reminiscent of the marginal-Fermi liquid
idea.  However, observation of a fractional
exponent (less than unity) in the power-law fall-off is more
consistent with the hidden-FL idea.

(ii) similar anomalous behavior is seen in optical conductivity
studies in near-optimally doped cuprates as well
as YbRh$_{2}$Si$_{2}$.

 (i) and (ii) show that {\it both} spin and charge correlations are
singular in a very anomalous sense in the
``strange'' metal.  To the extent that we are unaware of a calculation
where (i) and (ii) follow from analysis of
a single model, it is of obvious interest to consider this issue
within our approach.  We now study these issues.

 The fact that the ``RKKY'' interaction between localized ($a$-band)
moments is now mediated by critical fermionic
degrees of freedom, as discussed above, has remarkable consequences
for magnetic fluctuations.  A localized spin
is now ``Kondo'' coupled to these critical fermions.  Alternatively, a
local moment experiences a dynamically
fluctuating ``critical magnetic field'' due to the power-law form of
$G_{bb}(\omega)$.  Explicitly, this coupling is
$H'\simeq J_{K}\sum_{i}{\bf S}_{i,a}.{\bf \sigma}_{i,b} \simeq
J_{K}\sum_{i}{\bf S}_{i,a}.{\bf h}_{i,a}$, where the
``field'', ${\bf h}_{i,a}={\bf h}_{i,a}(\tau)$ encodes the singular
nature of the $b$-fermion spectrum found in DMFT.
 The effective superexchange generated between neighboring localized
moments is thus related to the temporal
correlations of this ``fluctuating field''.  This can now be treated
at an effective level as a problem of
a localized spin system
coupled to singular ``magnetic field'' fluctuations~\cite{sen}.  In
our case, the exponent in the singular
fluctuations of the ``magnetic field'' now enter from the singular
nature of $G_{bb}(\omega)$ found above.
As shown for this simplified problem, the spin-spin correlator now
reflects the critical power-law spectrum of the
``conduction sea'' via coupling to a ``critical'' bath.  In imaginary
time, the spin-spin correlator reads
$\chi_{ss}^{a}(\tau-\tau')=J_{K}^{2}\chi_{ss}^{(b)}(\tau-\tau')\simeq
(\tau-\tau')^{-(2-\mu)}$, where $\chi_{ss}^{(b)}(\tau)$ is the
dynamicakl spin susceptibility of the ``critical bath''.  Extracting
the functional form of $\mu$ on
the exponent of the power-law behavior in $G_{bb}(\omega)$ requires
more work.  We leave this detail for future work,
 in keeping with our aim here, which is to qualitatively show how the
anomalous magnetic fluctuations are intimately
linked to branch cut feature(s) in the one-fermion spectral functions.
Extending this to finite $T$, Fourier
transforming to energy variables, and analytically continuing the
imaginary (Matsubara) frequencies on to the real
energy axis yields the dynamical spin susceptibility that can be
compared to inelastic neutron scattering results.        The dynamical
susceptibility is~\cite{sensa}

\be
\chi({\bf q},\omega)=A({\bf q})T^{-(1-\mu)}F(\omega/T)
\ee,
where $F(y)=y^{1-\mu}$sinh$(y/2)|\Gamma(\frac{\mu}{2}-\frac{iy}{2})|^{2}$,
i.e, it shows the famous $\omega/T$ scaling.
 In the heavy LFL state, having a sharp LFL quasiparticle peak in both
$G_{aa},G_{bb}$ now implies that the
spin degrees of freedom experience a fluctuating magnetic field as
well, but one whose correlations are of the
usual LFL type.  Namely, the ``magnetic field'', or the $b$-fermion
spin correlator now goes off like
$(\tau-\tau')^{-2}$.  Hence, the spin correlation function now shows
conventional long-time behavior,
$\chi_{ss}^{a}(\tau-\tau')=J_{K}^{2}\chi_{ss}^{(b)}(\tau-\tau')\simeq
(\tau-\tau')^{-2}$, characteristic of a LFL,
as it must.

 We emphasise that a ``locally critical'' singular magnetic
fluctuation spectrum is generated in our case by the
{\it local} ``Kondo'' coupling of a local moment (arising from Mott
localization of $a$-fermions) to a critical
fermionic ``bath'' (via the infra-red singular G$_{bb}(\omega)$) in
the DMFT above.  This is physically very
different from how a similar effect arises in the Kondo lattice
model~\cite{si}: there, it arises as a competition
between the local (usual) Kondo effect and the RKKY intersite
interactions, precisely at the point where both heavy
LFL and magnetic ordering scales are simultaneously suppressed at
$T=0$: it is thus necessarily a quantum
criticsl {\it point}.  In our case, the ``strange'' metal arises
as a phase for a finite range of
$U_{fc}^{(1)}=0.6$~eV$<U_{fc}<U_{fc}^{(2)}$.  In our case, however,
there is
no need that local-critical magnetic fluctuations be tied to AF order:
we only require local coupling of local
moments to a critical fermionic bath, and the latter arises in a
selective-Mott scenario, as we have found here.
Thus, our results are again in accord with the FL$^{*}$ idea, which
does not {\it require} the strange metal
features to be tied down to onset of magnetic order.  While this
may be relevant to recent experiments on Co- and Ir-doped
YbRh$_{2}$Si$_{2}$~\cite{brando}, where it is found
that the ``strange'' transition (denoted by $T^{*}(p)$) and the AF
line ($T_{N}(p)$) do {\it not} coincide in
general as $T\rightarrow 0$, it is a non-trivial issue in this
specific context, possibly involving quantum
frustration~\cite{piers}, and is out of scope of the present work.

\begin{figure}
{\includegraphics[angle=270,width=\columnwidth]{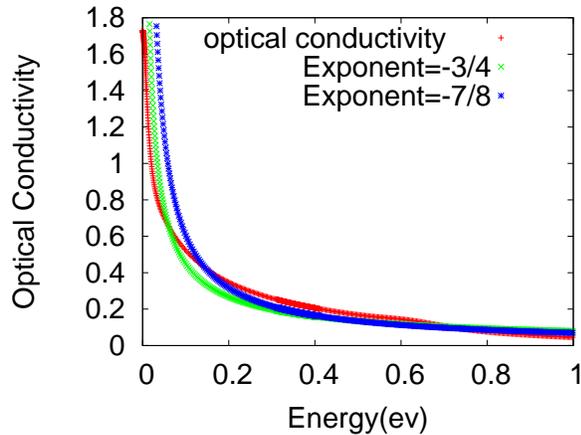}
}

\caption{(Color online) Optical conductivity in the critical metal
($u=3, \, u^{'}=1.6$, other parameters same as fig.1)
, clearly
showing $\omega^{-\nu}$ behavior with $\nu\simeq 3/4$ at low energy, in strong
 contrast to a smeared lorentzian expected for any heavy LFL state.
}
\label{fig5}
\end{figure}

 Finally, it is also of interest to study the optical conductivity.
It has been widely used with success to
study correlated systems, and shows unique, ill-understood features in
near-optimally doped cuprates and
YbRh$_{2}$Si$_{2}$.
 We show the full optical conductivity in the QC phase in
Fig.~\ref{fig5}.  In DMFT, this involves only the full
DMFT propagators, as vertex corrections rigorously drop out due to the
locality of the self-energy, or are small
enough to be negligible in multi-orbital systems~\cite{silke}.
Physically, the ``irrelevance'' of vertex corrections
in the strange metal can be argued by appeal to DMFT results.  In the
non-LFL metal, single-electron(hole)-like
quasiparticles do not live long enough to effectively interact with
each other in the ``intermediate'' state of the
(electron-hole) scattering process; i.e, that the soft multi-particle
excitations represented by the incoherent
(singular branch-cut) continuum in the DMFT results are scattered
before they can recohere into LFL quasiparticles.
To the extent that these processes are precisely the ones entering the
irreducible vertex in the Bethe-Salpeter equation for the
conductivity, their ineffectiveness in the above sense
implies that it is a good approximation to ignore vertex corrections.
Actually, a similar argument, referred to as
``holon non-drag'' regime in tomographic Luttinger liquid and
hidden-FL theories, was previously used for neglecting vertex
corrections in the
same context~\cite{phil}.
With these methodological clarifications, we find, remarkably, that
$\sigma(\omega) \simeq \omega^{-\nu}$ with $\nu\simeq3/4$ up to rather
high energy $O(1.0)$~eV, in sharp contrast to the $\omega^{-2}$
form expected for any LFL state.  Following Van der Marel et
al.~\cite{yrsi}, we estimate that
the transport scattering rate,
$\Gamma(\omega)\simeq \omega$cot$(\pi\nu/2)$, {\it linear} in $\omega$, while
the dynamical effective mass, $m^{*}(\omega)/m\simeq
\omega^{-(1-\nu)}$.  Such an unusual optical response is one
of the benchmarks of the ``strange'' metal, and is seen in optimally
doped cuprates and $YbRh_{2}Si_{2}$~\cite{yrsi}.

Thus, our DMFT based selective-Mott scenario yields a consistent
description of the ``strange'' features in
{\it both} charge and spin fluctuation channels in a single framework.

\section{Instabilities of the ``Strange'' Metal}

 Finally, the local critical metal we find has a large degeneracy
(finite entropy per site) and so can only be a
stable state of matter at low but intermediate $T$.  It must
eventually give way to either a heavy LFL, as described
above, or to a multiplicity of (competing) ordered states.
Irrelevance of one-electron mixing in our case is reminiscent of
coupled $D=1$ Luttinger liquids~\cite{phil}, where
irrelevance of one-particle inter-chain hopping favors two-particle
{\it inter-chain} coherence and ordered states.  In our case, in
analogy with the $D=1$ case, irrelevance of $V_{fc}(k)$ in the QC
metal
favors {\it intersite} and inter-orbital two-particle coherence via
$H_{res}\simeq (V_{fc}^{2}/U_{fc})\sum_{<i,j>,\sigma,\sigma'}(a_{i\sigma}^{\dag}b_{j\sigma}b_{j\sigma'}^{\dag}a_{i\sigma'}+h.c)$.
 Within DMFT, an exact (Hartree-Fock)
decoupling of $H_{res}$ gives two mutually {\it competing} p-h
and p-p order parameters, $\Delta_{ab}^{(1)}=\langle
a_{i\sigma}^{\dag}b_{j\sigma}\rangle$ and $\Delta_{ab}^{(2)}=\langle
a_{i\sigma}^{\dag}b_{j,-\sigma}^{\dag}\rangle$, describing p-h
and p-p order parameters, $\Delta_{ab}^{(1)}=\langle
a_{i\sigma}^{\dag}b_{j\sigma}\rangle$ and $\Delta_{ab}^{(2)}=\langle
a_{i\sigma}^{\dag}b_{j,-\sigma}^{\dag}\rangle$, describing
unconventional excitonic and superconductive orders
(with $V_{fc}(k)\simeq($cos$k_{x}$-cos$k_{y})$, these are $d$-wave ordered
states at $T=0$).  In general, the type and symmetry of the ordered states
resulting from the QC metal will thus be dictated
by the specific ${\bf k}$-dependent form-factor of the residual interactions.
This is radically different from the normal state(s) and instabilities
of a weakly correlated LFL, which crucially depend on band-nesting
and/or saddle point features in the LDA band structures, and {\it
cannot}, by construction,
lead to power-law singularities in $D>1$.

\section{Conclusion}

 In conclusion, we have shown that a quantum critical metallic {\it phase},
originating from selective-Mott transition via the OC, is an
intrinsic property of the EPAM, at least within DMFT.  Both, the famed
$\omega/T$ singularities in response functions {\it and} the FS
reconstruction, now understood as a {\it quantum} Lifshitz transition,
naturally fall out as consequences of selective-Mott physics.  This critical
metallic phase is found to either revert back to a heavy LFL as
hybridization is tuned to relevance, or argued to give way to
(unconventional) competing
orders via direct two-particle instabilities.  Our findings are in close
accord with the hidden-FL and FL$^{*}$ theories, and hold promise for
understanding the ``strange'' metal phases in quantum matter.

\end{document}